\documentclass[pra,showpacs,superscriptaddress]{revtex4}

\begin{document}
\newcommand{\Prob}{\,\text{Prob}\,}
\newcommand{\tr}{\,\text{tr}}
\newcommand{\trd}{\,\text{tr}_d}
\newcommand{\sech}{\,\text{sech}}

\title{Quantum nonlinear dynamics of continuously measured systems}

\author{A.~J.~Scott}
\email{ajs@maths.uq.edu.au}
\affiliation{Department of Mathematics,}
\author{G.~J.~Milburn}
\email{milburn@physics.uq.edu.au}
\affiliation{Department of Physics, \\The University of Queensland, QLD
4072 Australia.}

\begin{abstract}
Classical dynamics is formulated as a Hamiltonian flow on phase space, while quantum mechanics is formulated as a unitary dynamics in Hilbert space. These different formulations have made it difficult to directly compare quantum and classical nonlinear dynamics. Previous solutions have focussed on computing quantities associated with a statistical ensemble such as variance or entropy. However a more direct comparison would compare classical predictions to the quantum for continuous simultaneous measurement of position and momentum of a single system. In this paper we give a theory of such measurement and show that chaotic behaviour in classical systems can be reproduced by continuously measured quantum systems.
\end{abstract}

\maketitle

\section{Introduction}
The Hamiltonian formulation of classical mechanics assigns simultaneous, arbitrarily accurate, values for the canonically conjugate position and momentum to distinguishable particles. Indeed in classical mechanics these simultaneous values are regarded as properties of the particles themselves; measurement simply reveals these values and need not, in principle, add uncertainty to their determination. In quantum mechanics the conjugate position and momentum are represented by noncommuting operators, $\hat{x}$ and $\hat{p}$, with $[\hat{x},\hat{p}]=i\hbar$. It follows that there is no physical state for which position and momentum can take dispersion free values. This is the content of the standard formulation of the uncertainty principle $\langle(\Delta\hat{x})^2\rangle\langle(\Delta\hat{p})^2\rangle\geq\hbar^2/4$, where $\Delta\hat{A}=\hat{A}-\langle\hat{A}\rangle$. This does not mean, however, that simultaneous measurements of position and momentum are impossible, only that such simultaneous measurements cannot be made arbitrarily accurate \cite{Busch85}.

Our objective is to show that the {\it observed} phase space, reconstructed from the classical stochastic measurement record of continuous joint measurements of position and momentum, corresponds to the classical phase space description with added noise. By a careful specification of the measurement model, including the relevant states of the apparatus, together with a Markov assumption, we show that the observed classical stochastic measurement record corresponds to the conditional quantum averages $\langle \hat{x}(t)\rangle$ and $\langle \hat{p}(t)\rangle$. We obtain a stochastic Schr\"{o}dinger equation for the conditional state of the system with complex noise, together with classical stochastic equations for the variables corresponding to the observed measurement records. This extends the results of Bhattacharya et al. \cite{habib} where only a position measurement was considered. Other authors \cite{gisin0,gisin,gisin2,spiller,schack,brun} have considered a general type of stochastic Schr\"{o}dinger equation (which may be interpreted as corresponding to a measurement) in the framework of {\it quantum state diffusion} to examine dissipative chaos in open systems. 

A model for the simultaneous measurement of position and momentum was given long ago by Arthurs and Kelly \cite{arthurs}. This model consists of two meters which are allowed to interact instantaneously with the system. The interaction couples one of the meters to position and the other to momentum, encoding the results of the measurement in the final states of the meters. Projective measurements are then made on each of the meter states separately. In more general terms the Arthurs-Kelly model is an explicit example of the Gelfand-Naimark-Segal theorem \cite{Conway} which enables us to write a positive joint probability density of two real variables (the actual measurement outcomes) in terms of an inner product on an extended Hilbert space (extended to include the two apparatus as well as the system to be measured). It can also be considered as an example of a {\it positive operator valued measure} \cite{POVM1,POVM2}. The Arthurs-Kelly model suggests a route to a phase space description of quantum mechanics based on the joint outcomes of simultaneous measurements of position and momentum. However, to achieve a phase space description of the dynamics it is necessary to consider a time sequence of such measurements. The Arthurs-Kelly model forces the conditional state of the system into a coherent state after a single measurement and is too strong for this purpose.

We generalize the Arthurs-Kelly model to allow for a weakening of the measurement. In this way repeated measurements can take place without invariably reducing the system to a coherent state. By simultaneously increasing the number of measurements and weakening the strength of each measurement, the continuous limit is achieved. We show that under continuous observation the evolution of the system state is described by an It\^{o} stochastic Schr\"{o}dinger equation. Unlike evolution via the ordinary Schr\"{o}dinger equation, continuous observation forces the conditional state of the system to remain localized. Hence the quantum mean of the phase space variables can be thought of as a trajectory. It is in this light that we investigate quantum nonlinear dynamics under continuous observation.

The theory of continuous measurement in quantum mechanics was first developed by Barchielli and co-workers \cite{barchielli1,barchielli2,barchielli3} (see also \cite{diosi,belavkin}) and some of the results in this paper are contained in their work. However our approach uses the Arthurs-Kelly measurement model as a starting point and follows the work on continuous position measurements by Caves and one of us in \cite{caves,milburn}. In this way our results have a clear physical interpretation.

In section II we generalize the Arthurs-Kelly measurement model. In section III we derive the stochastic Schr\"{o}dinger equation for continuous measurement which is then investigated analytically in section IV and simulated numerically in section V for a chaotic system. Finally, in section VI we discuss our results.

\section{Phase space measurements}
Following Braunstein et al. \cite{braunstein} we rederive and generalize the result implicit in the work of Arthurs and Kelly \cite{arthurs}. The simultaneous measurement of position and momentum of a one dimensional quantum system is achieved by constructing an interaction governed by the Hamiltonian
\begin{equation}
\hat{H}_T=\hat{H}(\hat{x},\hat{p})+\left(\textstyle\frac{1}{s}\hat{x}\hat{p}_1+s\hat{p}\hat{p}_2\right)\delta(t-t_r)
\end{equation}
which couples the system Hamiltonian $\hat{H}$, with two detectors $d_1$ and $d_2$. The measurement model also requires that the detectors have been prepared in the initial states
\begin{eqnarray}
\langle x_i|d_i\rangle &=& (\pi\hbar\Delta_i)^{-\frac{1}{4}}\exp\left(\frac{-x_i^2}{2\hbar\Delta_i}\right) \label{dxi} \\
\langle p_i|d_i\rangle &=& \left(\frac{\pi\hbar}{\Delta_i}\right)^{-\frac{1}{4}}\exp\left(\frac{-p_i^2\Delta_i}{2\hbar}\right) \label{dpi}
\end{eqnarray}
where $\Delta_1=\frac{\sigma^2}{2s}$ and $\Delta_2=\frac{s\sigma^2}{2}$. The parameter $s$ is called the {\it squeezing parameter} and biases the coupling so that one may obtain more information on either position or momentum. The parameter $\sigma$ will be used to weaken the measurement, decreasing the amount of information collected on both position and momentum.

Before the interaction the system state $\hat{\rho}$ and the combined detector state
\begin{equation}
\hat{\rho}_d=|d_1d_2\rangle\langle d_1d_2|\equiv|d_1\rangle\langle d_1|\otimes|d_2\rangle\langle d_2|
\end{equation}
are assumed to be uncorrelated with an initial combined density operator of $\hat{\rho}\otimes\hat{\rho}_d$. At $t=t_r$ the evolution operator for the interaction
\begin{equation}
\hat{U}_I\equiv\exp\left[-\textstyle\frac{i}{\hbar}\left(\textstyle\frac{1}{s}\hat{x}\hat{p}_1+s\hat{p}\hat{p}_2\right)\right]
\end{equation}
couples the system to the measurement apparatus, entangling the system and detector states. After the interaction the probability of finding the detector positions in the small area $[x_1,x_1+dx_1]\times[x_2,x_2+dx_2]$ is
\begin{eqnarray}
\Prob(x_1,x_2)dx_1dx_2 &=& \tr\trd\left(\hat{U}_I\hat{\rho}\hat{\rho}_d\hat{U}_I^\dag|x_1x_2\rangle\langle x_1x_2|\right)dx_1dx_2 \\
&=& \tr\left(\hat{\Upsilon}_{\!\sigma}(x_1,x_2)\hat{\rho}{\hat{\Upsilon}_{\!\sigma}(x_1,x_2)}^\dag\right)dx_1dx_2
\end{eqnarray}
where $\tr(\dots)$ and $\trd(\dots)$ are the traces over the system and detector states, respectively, $\Prob(x_1,x_2)$ is the probability density, and the {\it resolution operator}
\begin{eqnarray}
\hat{\Upsilon}_{\!\sigma}(x_1,x_2) &\equiv& \langle x_1x_2|\hat{U}_I|d_1d_2\rangle \\
&=& (2\pi\hbar)^{-1}\int dp_1dp_2\exp\left[-\textstyle\frac{i}{\hbar}\left(p_1\left(\textstyle\frac{1}{s}\hat{x}-x_1\right)+p_2\left(s\hat{p}-x_2\right)\right)\right]\langle p_1p_2|d_1d_2\rangle \\
&=& \hat{D}(sx_1,\textstyle\frac{1}{s}x_2)\hat{\Upsilon}_{\!\sigma}(0,0)\hat{D}(sx_1,\textstyle\frac{1}{s}x_2)^\dag.
\end{eqnarray}
The displacement operator
\begin{equation}
\hat{D}(\mu,\nu)\equiv\exp\left[-\textstyle\frac{i}{\hbar}\left(\mu\hat{p}-\nu\hat{x}\right)\right]
\end{equation}
and
\begin{equation}
\hat{\Upsilon}_{\!\sigma}(0,0)=(2\pi\hbar)^{-1}\int dp_1dp_2\langle p_1,-p_2|d_1d_2\rangle\hat{D}(sp_2,\textstyle\frac{1}{s}p_1)^\dag.
\label{Up00}
\end{equation}

We now define the annihilation operator
\begin{equation}
\hat{a}\equiv\textstyle\frac{1}{\sqrt{2\hbar}}\left(\sqrt{s}\hat{x}+i\textstyle\frac{1}{\sqrt{s}}\hat{p}\right)
\end{equation}
which satisfies
\begin{equation}
\left[\hat{a},\hat{a}^\dag\right]=1
\end{equation}
and rewrite the displacement operator as
\begin{equation}
\hat{D}(\zeta)\equiv\exp\left(\zeta\hat{a}^\dag-\zeta^*\hat{a}\right)
\end{equation}
where
$\zeta=\textstyle\frac{1}{\sqrt{2\hbar}}\left(\sqrt{s}\mu+i\textstyle\frac{1}{\sqrt{s}}\nu\right)$. It can be easily shown that
\begin{eqnarray}
&\hat{D}(\zeta)^\dag=\hat{D}(-\zeta)=\hat{D}(\zeta)^{-1}& \\
&\hat{D}(\zeta)\hat{a}\hat{D}(\zeta)^\dag=\hat{a}-\zeta.&
\end{eqnarray}
The coherent states \cite{perelomov} are now defined by applying the displacement operator onto the vacuum state
\begin{equation}
|\alpha\rangle\equiv\hat{D}(\alpha)|0\rangle.
\end{equation}
They satisfy the following relations
\begin{eqnarray}
\hat{a}|\alpha\rangle &&= \alpha|\alpha\rangle \label{a3}\\
\hat{D}(\beta)|\alpha\rangle &&=\exp\left[\textstyle\frac{1}{2}\left(\alpha^*\beta-\alpha\beta^*\right)\right]|\alpha+\beta\rangle \label{a4}\\
\langle\alpha|\beta\rangle &&=\exp\left(-\textstyle\frac{1}{2}|\alpha|^2-\textstyle\frac{1}{2}|\beta|^2+\alpha^*\beta\right) \label{a5}\\
\pi^{-1}\int d^2\alpha&&|\alpha\rangle\langle\alpha|=1. \label{a6}
\end{eqnarray}
Now defining
$\varrho\equiv\textstyle\frac{1}{\sqrt{2\hbar}}\left(\sqrt{s}p_2+i\textstyle\frac{1}{\sqrt{s}}p_1\right)$ we can rewrite (\ref{Up00}) as
\begin{equation}
\hat{\Upsilon}_{\!\sigma}(0,0)=(2\pi\hbar)^{-\frac{1}{2}}\sigma\pi^{-1}\int d^2\varrho\langle \sigma\varrho|0\rangle\hat{D}(\varrho)^\dag.
\end{equation}
Using the above relations and integrating we obtain
\begin{eqnarray}
\langle\alpha|\hat{\Upsilon}_{\!\sigma}(0,0)|\beta\rangle &=& (2\pi\hbar)^{-\frac{1}{2}}\textstyle\frac{2\sigma}{\sigma^2+1}\exp\left(-\textstyle\frac{1}{2}|\alpha|^2-\textstyle\frac{1}{2}|\beta|^2+\textstyle\frac{\sigma^2-1}{\sigma^2+1}\alpha^*\beta\right)\\
&=& (2\pi\hbar)^{-\frac{1}{2}}\textstyle\frac{2\sigma}{\sigma^2+1}\langle\alpha|\beta\rangle\exp\left(-\textstyle\frac{2}{\sigma^2+1}\alpha^*\beta\right)\\
&=& \langle\alpha|(2\pi\hbar)^{-\frac{1}{2}}\textstyle\frac{2\sigma}{\sigma^2+1}:\exp\left(-\textstyle\frac{2}{\sigma^2+1}\hat{a}^\dag\hat{a}\right):|\beta\rangle
\end{eqnarray}
where $:\cdot:$ denotes normal ordering. Hence,
\begin{equation}
\hat{\Upsilon}_{\!\sigma}(0,0)=(2\pi\hbar)^{-\frac{1}{2}}\textstyle\frac{2\sigma}{\sigma^2+1}:\exp\left(-\textstyle\frac{2}{\sigma^2+1}\hat{a}^\dag\hat{a}\right):.
\end{equation}
If we now define
$\chi=\chi_1+i\chi_2\equiv\textstyle\frac{1}{\sqrt{2\hbar}}\left(\sqrt{s}x_1+i\textstyle\frac{1}{\sqrt{s}}x_2\right)$ then the resolution operator becomes
\begin{eqnarray}
\hat{\Upsilon}_{\!\sigma}(x_1,x_2) &=& \hat{D}(\chi)\hat{\Upsilon}_{\!\sigma}(0,0)\hat{D}(\chi)^\dag \\
&=&
(2\pi\hbar)^{-\frac{1}{2}}\textstyle\frac{2\sigma}{\sigma^2+1}:\exp\left[-\textstyle\frac{2}{\sigma^2+1}(\hat{a}^\dag-\chi^*)(\hat{a}-\chi)\right]: \\
&\equiv& (2\hbar)^{-\frac{1}{2}}\hat{\Upsilon}_{\!\sigma}(\chi). \label{res}
\end{eqnarray}
The probability of finding the detector positions in the small area $[\chi_1,\chi_1+d\chi_1]\times[\chi_2,\chi_2+d\chi_2]$ is now
\begin{eqnarray}
\Prob(\chi)d^2\chi &=& 2\hbar\Prob(x_1,x_2)d^2\chi \\
&=&
\tr\left(\hat{\Upsilon}_{\!\sigma}(\chi)\hat{\rho}{\hat{\Upsilon}_{\!\sigma}(\chi)}^\dag\right)d^2\chi \\
&=& \tr\left(\hat{F}_{\!\sigma}(\chi)\hat{\rho}\right)d^2\chi
\end{eqnarray}
where $d^2\chi=d\chi_1d\chi_2$ and
\begin{eqnarray}
\hat{F}_{\!\sigma}(\chi) &\equiv& {\hat{\Upsilon}_{\!\sigma}(\chi)}^\dag\hat{\Upsilon}_{\!\sigma}(\chi) \\
&=&
\pi^{-1}\left(\textstyle\frac{2\sigma}{\sigma^2+1}\right)^2:\exp\left[-\left(\textstyle\frac{2\sigma}{\sigma^2+1}\right)^2(\hat{a}^\dag-\chi^*)(\hat{a}-\chi)\right]:
\end{eqnarray}
is an {\it effect density} \cite{POVM1}. It can be easily shown that
\begin{eqnarray}
\int\chi^n\hat{F}_{\!\sigma}(\chi)d^2\chi &=& \hat{a}^n \\
\int|\chi|^2\hat{F}_{\!\sigma}(\chi)d^2\chi &=& \hat{a}\hat{a}^\dag+\left(\textstyle\frac{\sigma^2-1}{2\sigma}\right)^2.
\end{eqnarray}
Hence, defining the notion of a mean for this measurement process
\begin{eqnarray}
\langle f(\chi)\rangle_\sigma &\equiv& \int f(\chi)\Prob(\chi)d^2\chi \\
&=& \int f(\chi)\tr\left(\hat{F}_{\!\sigma}(\chi)\hat{\rho}\right)d^2\chi
\end{eqnarray}
we find that
\begin{eqnarray}
\langle\chi\rangle_\sigma &=& \langle\hat{a}\rangle \label{m1}\\
\langle|\chi|^2\rangle_\sigma &=& \langle\hat{a}\hat{a}^\dag\rangle+\left(\textstyle\frac{\sigma^2-1}{2\sigma}\right)^2 \label{m2}
\end{eqnarray}
or
\begin{eqnarray}
\langle x_1\rangle_\sigma=\langle\hat{x}\rangle, \quad&&\quad \langle x_2\rangle_\sigma=\langle\hat{p}\rangle \\
\langle {x_1}^2\rangle_\sigma=\langle\hat{x}^2\rangle&&+\textstyle\frac{\hbar}{s}\left(\textstyle\frac{1+\sigma^4}{4\sigma^2}\right) \\
\langle {x_2}^2\rangle_\sigma=\langle\hat{p}^2\rangle&&+\hbar s\left(\textstyle\frac{1+\sigma^4}{4\sigma^2}\right)
\end{eqnarray}
where $\langle\hat{A}\rangle=\tr(\hat{A}\hat{\rho})$ is the quantum expectation. Thus the readout variables $x_1$ and $x_2$ give, respectively, the position and momentum of the system with additional noise dependent on $\sigma$. We obtain maximal information from the system when the variances are at a minimum. That is, when $\sigma=1$. In this case
\begin{eqnarray}
\hat{F}_{\! 1}(\chi) &=& \pi^{-1}:\exp\left[-(\hat{a}^\dag-\chi^*)(\hat{a}-\chi)\right]: \\
&=& \pi^{-1}|\chi\rangle\langle\chi|
\end{eqnarray}
and the probability density reduces to the Husimi density or $Q$ function
\begin{eqnarray}
\Prob(\chi) &=& \pi^{-1}\langle\chi|\hat{\rho}|\chi\rangle \\
&=& \pi^{-1}Q(\chi).
\end{eqnarray}
Suppose we take a measurement and obtain the outcome $\chi'$. As a consequence of the strength of this measurement the system state collapses to
\begin{eqnarray}
\hat{\rho}'&=&\frac{1}{\Prob(\chi')}\hat{\Upsilon}_{\! 1}(\chi')\hat{\rho}{\hat{\Upsilon}_{\! 1}(\chi')}^\dag\\
&=&\frac{1}{\langle\chi'|\hat{\rho}|\chi'\rangle}|\chi'\rangle\langle\chi'|\hat{\rho}|\chi'\rangle\langle\chi'| \\
&=& |\chi'\rangle\langle\chi'|.
\end{eqnarray}
However when $\sigma\gg 1$ the resolution operator has the asymptotic expansion
\begin{eqnarray}
\hat{\Upsilon}_{\!\sigma}(\chi)&=&\pi^{-\frac{1}{2}}\textstyle\frac{2}{\sigma}\Bigl[1-\textstyle\frac{1}{\sigma^2}\left(1+2(\hat{a}^\dag-\chi^*)(\hat{a}-\chi)\right)\Bigr]+O(\textstyle\frac{1}{\sigma^5}). \label{asympx}
\end{eqnarray}
Using this result one can show that the system state conditioned on the measurement outcome $\chi'$ has the expansion
\begin{equation}
\hat{\rho}'=\hat{\rho}+\textstyle\frac{2}{\sigma^2}\Bigl\{\tr\Bigl[(\hat{a}^\dag-\chi'^*)(\hat{a}-\chi')\hat{\rho}\Bigr]-(\hat{a}^\dag-\chi'^*)(\hat{a}-\chi'),\hat{\rho}\Bigr\}+O(\textstyle\frac{1}{\sigma^4})
\end{equation}
and thus, if $\sigma$ is large enough, the process of measurement will have negligible effect on the system.

\section{Continuous measurement}

Consider a sequence of phase space measurements governed by the Hamiltonian
\begin{equation}
\hat{H}_T=\hat{H}(\hat{x},\hat{p})+\left(\textstyle\frac{1}{s}\hat{x}\hat{p}_1+s\hat{p}\hat{p}_2\right)\sum_{n=0}^\infty\delta(t-n{\delta t})
\end{equation}
where after each measurement the detectors are reset into the initial states given by equation (\ref{dxi}) or (\ref{dpi}). The assumption that the detectors are reset is equivalent to making a Markov assumption for a single apparatus coupled to the system. By resetting the detector at each time step we ensure that no coherent memory of the system state survives in the states of the apparatus. Following \cite{caves} we will first derive the master equation for unconditional (or nonselective) evolution of the system density operator in the continuous limit ${\delta t}\rightarrow 0$, $\sigma\rightarrow\infty$. By unconditional evolution we mean that no account is taken of the measured results. Thus after each measurement occurs we ignore the result and average over all possible measurement outcomes. If we denote the system density operator immediately before the $n$-th measurement by $\hat{\rho}(n{\delta t})$ then
\begin{equation}
\hat{\rho}(n{\delta t}+{\delta t})= \hat{U}\int d^2\chi\hat{\Upsilon}_{\!\sigma}(\chi)\hat{\rho}(n{\delta t}){\hat{\Upsilon}_{\!\sigma}(\chi)}^\dag{\hat{U}}^\dag
\end{equation}
where
\begin{eqnarray}
\hat{U} &\equiv& \exp\left(-\textstyle\frac{i}{\hbar}\hat{H}{\delta t}\right) \label{u}\\
&=& 1-\textstyle\frac{i}{\hbar}\hat{H}{\delta t}+O({\delta t}^2).
\label{asympu}
\end{eqnarray}
For any operator $\hat{A}$ it is possible to show that
\begin{equation}
\int d^2\chi\hat{\Upsilon}_{\!\sigma}(\chi)\hat{A}{\hat{\Upsilon}_{\!\sigma}(\chi)}^\dag=\hat{A}-\textstyle\frac{1}{\sigma^2}\left[\hat{a},\left[\hat{a}^\dag,\hat{A}\right]\right]+O(\textstyle\frac{1}{\sigma^4}).
\end{equation}
Hence we obtain
\begin{equation}
\frac{\hat{\rho}(n{\delta t}+{\delta t})-\hat{\rho}(n{\delta t})}{{\delta t}}=-\textstyle\frac{i}{\hbar}\left[\hat{H},\hat{\rho}(n{\delta t})\right]-\textstyle\frac{1}{{\delta t}\sigma^2}\left[\hat{a},\left[\hat{a}^\dag,\hat{\rho}(n{\delta t})\right]\right]+O({\delta t})+O(\textstyle\frac{1}{\sigma^2})+O(\textstyle\frac{1}{{\delta t}\sigma^4}).
\end{equation}
By setting $t=n{\delta t}$ and taking the continuous limit ${\delta t}\rightarrow 0$, $\sigma\rightarrow\infty$, with $\gamma=1/{\delta t}\sigma^2$ held constant, we obtain the master equation for unconditional evolution:
\begin{eqnarray}
\frac{d\hat{\rho}}{dt} &=&
-\textstyle\frac{i}{\hbar}\left[\hat{H},\hat{\rho}\right]-\gamma\left[\hat{a},\left[\hat{a}^\dag,\hat{\rho}\right]\right] \label{uncon}\\
 &=&
-\textstyle\frac{i}{\hbar}\left[\hat{H},\hat{\rho}\right]-\textstyle\frac{1}{2\hbar}\Gamma_1\Big[\hat{x},\Big[\hat{x},\hat{\rho}\Big]\Big]-\textstyle\frac{1}{2\hbar}\Gamma_2\Big[\hat{p},\Big[\hat{p},\hat{\rho}\Big]\Big]
\label{xpuncondequ}
\end{eqnarray}
where $\Gamma_1\equiv\gamma s$ and $\Gamma_2\equiv\textstyle\frac{\gamma}{s}$. This equation has already been derived by Barchielli et al. \cite{barchielli1}, but their approach is somewhat different. By setting $\Gamma_2=0$ in (\ref{xpuncondequ}) we obtain the unconditional master equation for continuous position measurements previously derived in \cite{caves}.

We now wish to derive the conditional (or selective) master equation for the system density operator. In this case the evolution of the system is conditioned on a history of measurement readouts
\begin{equation}
\left\{\chi(n{\delta t})\right\}=\left\{\chi(0),\chi({\delta t}),\chi(2{\delta t}),\dots\right\}
\end{equation}
where $\chi(n{\delta t})$ is the detector position for the $n$-th measurement. Hence, if $\hat{\rho}(n{\delta t})$ is the system density operator immediately before the $n$-th measurement, then
\begin{equation}
\hat{\rho}(n{\delta t}+{\delta t})= \hat{U}\left[\Prob(\chi(n{\delta t}))\right]^{-1}\hat{\Upsilon}_{\!\sigma}(\chi(n{\delta t}))\hat{\rho}(n{\delta t}){\hat{\Upsilon}_{\!\sigma}(\chi(n{\delta t}))}^\dag{\hat{U}}^\dag.
\end{equation}
To proceed we extend the definition of the readout variable by setting
\begin{equation}
\chi(t)=\chi(n{\delta t}) \quad\text{ for }\quad n{\delta t}\leq t<(n+1){\delta t}
\end{equation}
and introduce the new variable
\begin{equation}
{\cal X}(t)\equiv\int_{0}^{t}\chi(t')dt'.
\end{equation}
Hence ${\cal X}(0)=0$ and
\begin{equation}
{\cal X}(n{\delta t})={\delta t}\sum_{m=0}^{n-1}\chi(m{\delta t}) \quad\text{ for }\quad n\geq 1.
\end{equation}
Using (\ref{m1}) and (\ref{m2}) we obtain
\begin{eqnarray}
E_c\left(\chi(n{\delta t})\right) &=& \langle\chi(n{\delta t})\rangle_\sigma \\
&=& \tr\left(\hat{a}\hat{\rho}(n{\delta t})\right) \label{mean}\\
V_c\left(\chi(n{\delta t})\right) &=& \left\langle\bigl|\chi(n{\delta t})-\langle\chi(n{\delta t})\rangle_{\sigma}\bigr|^2\right\rangle_{\sigma} \\
&=& \langle|\chi(n{\delta t})|^2\rangle_{\sigma}-\left|\langle\chi(n{\delta t})\rangle_{\sigma}\right|^2 \\
&=& \tr\left(\hat{a}\hat{a}^\dag\hat{\rho}(n{\delta t})\right)-\left|\tr\left(\hat{a}\hat{\rho}(n{\delta t})\right)\right|^2+\left(\textstyle\frac{\sigma^2-1}{2\sigma}\right)^2.
\label{var}
\end{eqnarray}
where the subscript $c$ has been added to emphasize that the mean and variance are conditioned through $\hat{\rho}$ on the entire history of measurement readouts. Now letting
\begin{equation}
\delta{\cal X}(n{\delta t})={\cal X}(n{\delta t}+{\delta t})-{\cal X}(n{\delta t})={\delta t}\chi(n{\delta t}) \label{Xincrement1}
\end{equation}
we find that
\begin{eqnarray}
E_c\left(\delta{\cal X}(n{\delta t})\right) &=& {\delta t}\tr\left(\hat{a}\hat{\rho}(n{\delta t})\right) \\
V_c\left(\delta{\cal X}(n{\delta t})\right) &=& {\delta t}^2\left(\tr\left(\hat{a}\hat{a}^\dag\hat{\rho}(n{\delta t})\right)-\left|\tr\left(\hat{a}\hat{\rho}(n{\delta t})\right)\right|^2+\left(\textstyle\frac{\sigma^2-1}{2\sigma}\right)^2\right) \\
&=& \textstyle\frac{1}{4}\gamma^{-1}{\delta t}+O({\delta t}^2).
\end{eqnarray}
where we have set $\gamma=1/{\delta t}\sigma^2$ to be constant in anticipation of the continuous limit. Hence for ${\delta t}$ small enough we can approximate increments in the variable ${\cal X}$ by
\begin{equation}
\delta{\cal X}(n{\delta t})=\tr\left(\hat{a}\hat{\rho}(n{\delta t})\right){\delta t}+ \textstyle\frac{1}{2}\gamma^{-\frac{1}{2}}\delta\xi(n{\delta t}) \label{Xincrement2}
\end{equation}
where it is understood that the complex It\^{o} increment $\delta \xi$ is of order ${\delta t}^\frac{1}{2}$ and satisfies $E\left(\delta\xi\right)=0$, $E\left(\delta\xi^*\delta \xi\right)={\delta t}$. In the continuous limit ${\delta t}\rightarrow 0$ with $t=n{\delta t}$ constant we have
\begin{equation}
d{\cal X}(t)=\tr\left(\hat{a}\hat{\rho}(t)\right)dt+\textstyle\frac{1}{2}\gamma^{-\frac{1}{2}}d\xi(t) \label{calX}
\end{equation}
where $\xi(t)$ is a complex Wiener process \cite{gardiner} and the It\^{o} differential $d\xi$ satisfies the algebra
\begin{eqnarray}
E\left(d\xi\right) &=& 0 \\
d\xi^*d\xi &=& dt \\
d\xi^2 &=& 0 \\
d\xi dt &=& 0.
\end{eqnarray}
To simplify the following we will always replace $\delta\xi^*\delta \xi$ by ${\delta t}$ and set $\delta \xi^2=\delta\xi^{*2}=0$ in anticipation of the above algebra in the continuous limit. Using (\ref{Xincrement1}) together with (\ref{Xincrement2}) we obtain the following asymptotic expansion for the resolution operator (\ref{res})
\begin{equation}
\hat{\Upsilon}_{\!\sigma}(\chi)=2\left(\textstyle\frac{\gamma{\delta t}}{\pi e}\right)^{\frac{1}{2}}\left(1+\gamma^\frac{1}{2}\left(\hat{{\cal A}}^\dag\delta \xi +\hat{{\cal A}}\delta\xi^*\right)-\gamma{\delta t}\left(\hat{{\cal A}}^\dag\hat{{\cal A}}+\textstyle\frac{1}{2}\right)\right)+O({\delta t}^2)
\end{equation}
where
\begin{equation}
\hat{{\cal A}}\equiv\hat{a}-\tr\left(\hat{a}\hat{\rho}\right)
\end{equation}
and it is understood that $\hat{\rho}=\hat{\rho}(n{\delta t})$, $\chi=\chi(n{\delta t})$ and $\delta \xi=\delta \xi(n{\delta t})$. Hence we find that
\begin{equation}
\hat{\Upsilon}_{\!\sigma}(\chi)\hat{\rho}{\hat{\Upsilon}_{\!\sigma}(\chi)}^\dag=4\textstyle\frac{\gamma{\delta t}}{\pi e}\left(\hat{\rho}+\gamma^\frac{1}{2}\left\{\hat{{\cal A}}^\dag\delta \xi+\hat{{\cal A}}\delta\xi^*,\hat{\rho}\right\}-\gamma{\delta t}\left[\hat{{\cal A}},\left[\hat{{\cal A}}^\dag,\hat{\rho}\right]\right]\right)+O({\delta t}^3)
\end{equation}
and thus, using (\ref{asympu}) we obtain
\begin{eqnarray}
\hat{\rho}(n{\delta t}+{\delta t}) &=& \hat{\rho}-\textstyle\frac{i{\delta t}}{\hbar}\left[\hat{H},\hat{\rho}\right]+\gamma^\frac{1}{2}\left\{\hat{{\cal A}}^\dag\delta \xi +\hat{{\cal A}}\delta\xi^*,\hat{\rho}\right\}-\gamma{\delta t}\left[\hat{{\cal A}},\left[\hat{{\cal A}}^\dag,\hat{\rho}\right]\right]+O({\delta t}^\frac{3}{2}) \\
&=& \hat{\rho}-\textstyle\frac{i{\delta t}}{\hbar}\left[\hat{H},\hat{\rho}\right]-\gamma{\delta t}\left[\hat{a},\left[\hat{a}^\dag,\hat{\rho}\right]\right]+\gamma^\frac{1}{2}{\cal H}[\hat{a}^\dag]\hat{\rho}\delta \xi+\gamma^\frac{1}{2}{\cal H}[\hat{a}]\hat{\rho}\delta\xi^*+O({\delta t}^\frac{3}{2})
\end{eqnarray}
where we have defined the superoperator
\begin{equation}
{\cal H}[\hat{A}]\hat{\rho}\equiv\left\{\hat{A}-\tr(\hat{A}\hat{\rho}),\hat{\rho}\right\}.
\end{equation}
In the limit ${\delta t}\rightarrow 0$ we obtain the master equation for conditional evolution:
\begin{eqnarray}
d\hat{\rho}(t) &=& \hat{\rho}(t+dt)-\hat{\rho}(t) \\
&=& -\textstyle\frac{i}{\hbar}\left[\hat{H},\hat{\rho}(t)\right]dt-\gamma\left[\hat{ a},\left[\hat{a}^\dag,\hat{\rho}(t)\right]\right]dt+\gamma^\frac{1}{2}{\cal H}[\hat{a}^\dag]\hat{\rho}(t)d\xi(t)+\gamma^\frac{1}{2}{\cal H}[\hat{a}]\hat{\rho}(t)d\xi(t)^*.
\end{eqnarray}
It is easy to see that upon averaging this stochastic differential equation we reproduce our original master equation for unconditional evolution (\ref{uncon}). However, note that unlike the unconditional equation, this equation preserves the pure state property of $\hat{\rho}$. One can easily prove this by showing $d\hat{\rho}^2=d\hat{\rho}$ under the assumption that $\hat{\rho}=|\psi\rangle\langle\psi|$, where $d\hat{\rho}^2\equiv\{d\hat{\rho},\hat{\rho}\}+d\hat{\rho}d\hat{\rho}$. As a consequence of this fact, the above master equation has an analogue for pure state evolution in terms of a stochastic Schr\"{o}dinger equation:
\begin{eqnarray}
d|\psi(t)\rangle =-\textstyle\frac{i}{\hbar}\hat{H}|\psi(t)\rangle dt-\gamma&&\left(\hat{a}^\dag\hat{a}+\textstyle\frac{1}{2}-\langle\hat{a}^\dag\rangle\hat{a}-\hat{a}^\dag\langle\hat{a}\rangle+|\langle\hat{a}\rangle|^2\right)|\psi(t)\rangle dt \nonumber \\
&&+\gamma^\frac{1}{2}\left(\hat{a}^\dag-\langle\hat{a}^\dag\rangle\right)|\psi(t)\rangle d\xi(t)+\gamma^\frac{1}{2}\Big(\hat{a}-\langle\hat{a}\rangle\Big)|\psi(t)\rangle d\xi(t)^* \label{stochschro}
\end{eqnarray}
where $\langle\hat{A}\rangle=\langle\psi(t)|\hat{A}|\psi(t)\rangle$. In terms of position and momentum variables the master equation for conditional evolution reads as
\begin{equation}
d\hat{\rho} = -\textstyle\frac{i}{\hbar}\left[\hat{H},\hat{\rho}\right]dt-\textstyle\frac{1}{2\hbar}\Gamma_1\Big[\hat{x},\Big[\hat{x},\hat{\rho}\Big]\Big]dt-\textstyle\frac{1}{2\hbar}\Gamma_2\Big[\hat{p},\Big[\hat{p},\hat{\rho}\Big]\Big]dt+\hbar^{-\frac{1}{2}}{\Gamma_1}^{\frac{1}{2}}{\cal H}[\hat{x}]\hat{\rho}dW_1+\hbar^{-\frac{1}{2}}{\Gamma_2}^{\frac{1}{2}}{\cal H}[\hat{p}]\hat{\rho}dW_2 \label{xpcondequ}
\end{equation}
where
\begin{eqnarray}
E\left(dW_i\right) &=& 0 \\
dW_idW_j &=& \delta_{ij}dt \\
\textstyle\frac{1}{\sqrt{2}}\left(dW_1+idW_2\right) &=& d\xi
\end{eqnarray}
and the readout variables $x_1$ and $x_2$ obey the following stochastic processes
\begin{eqnarray}
dX_1(t)=\tr\left(\hat{x}\hat{\rho}(t)\right)dt+ &&\textstyle\frac{1}{2}\hbar^\frac{1}{2}{\Gamma_1}^{-\frac{1}{2}}dW_1(t) \\
dX_2(t)=\tr\left(\hat{p}\hat{\rho}(t)\right)dt+ &&\textstyle\frac{1}{2}\hbar^\frac{1}{2}{\Gamma_2}^{-\frac{1}{2}}dW_2(t)
\end{eqnarray}
with
\begin{equation}
X_1(t)\equiv\int_{0}^{t}x_1(t')dt' \qquad X_2(t)\equiv\int_{0}^{t}x_2(t')dt' .
\end{equation}
By setting $\Gamma_2=0$ in (\ref{xpcondequ}) we obtain the conditional master equation for continuous position measurements previously derived in \cite{milburn}. Note that $x_1$ and $x_2$ are charged by stationary white noise
\begin{eqnarray}
x_1=\langle\hat{x}\rangle+ &&\textstyle\frac{1}{2}\hbar^\frac{1}{2}{\Gamma_1}^{-\frac{1}{2}}\dot{W_1} \label{xa}\\
x_2=\langle\hat{p}\rangle+ &&\textstyle\frac{1}{2}\hbar^\frac{1}{2}{\Gamma_2}^{-\frac{1}{2}}\dot{W_2} \label{xb}
\end{eqnarray}
making their graph highly irregular. It is thus better to represent the measured trajectory by $\langle\hat{x}\rangle$ and $\langle\hat{p}\rangle$.

\section{Analytical investigations}

We will now investigate the effect of measurement on the system state by setting $\hat{H}=0$ and defining
\begin{eqnarray}
V_x&\equiv&E\left(\langle\hat{x}^2\rangle-\langle\hat{x}\rangle^2\right) \\
V_p&\equiv&E\left(\langle\hat{p}^2\rangle-\langle\hat{p}\rangle^2\right) \\
C_{xp}&\equiv&E\Big(\textstyle\frac{1}{2}\langle\hat{x}\hat{p}\rangle+\textstyle\frac{1}{2}\langle\hat{p}\hat{x}\rangle-\langle\hat{x}\rangle\langle\hat{p}\rangle\Big)
\end{eqnarray}
where $V_x$ and $V_p$ are the expected variances in position and momentum, and $C_{xp}$ is the expected covariance between position and momentum. The average is taken over all possible measurement histories. One can then derive the following set of coupled differential equations:
\begin{eqnarray}
\frac{dV_x}{dt}&=&\frac{\gamma\hbar}{s}-\frac{4s\gamma}{\hbar}{V_x}^2-\frac{4\gamma}{s\hbar}{C_{xp}}^2 \\
\frac{dV_p}{dt}&=&s\gamma\hbar-\frac{4\gamma}{s\hbar}{V_p}^2-\frac{4s\gamma}{\hbar}{C_{xp}}^2 \\
\frac{dC_{xp}}{dt}&=&-\frac{4\gamma}{\hbar}C_{xp}\left(sV_x+\frac{1}{s}V_p\right).
\end{eqnarray}
The solutions of which are
\begin{eqnarray}
V_x(t)&=&\frac{\hbar}{2s}\frac{\Big(2sV_x^0+\hbar\tanh(2\gamma t)\Big)\Big(s\hbar+2V_p^0\tanh(2\gamma t)\Big)-4s{C_{xp}^0}^2\tanh(2\gamma t)}{\Big(\hbar+2sV_x^0\tanh(2\gamma t)\Big)\Big(s\hbar+2V_p^0\tanh(2\gamma t)\Big)-4s{C_{xp}^0}^2\tanh^2(2\gamma t)} \\
&\rightarrow&\frac{\hbar}{2s}\quad\text{ as }\quad t\rightarrow\infty
\nonumber\\
V_p(t)&=&\frac{s\hbar}{2}\frac{\Big(2V_p^0+s\hbar\tanh(2\gamma t)\Big)\Big(\hbar+2sV_x^0\tanh(2\gamma t)\Big)-4s{C_{xp}^0}^2\tanh(2\gamma t)}{\Big(\hbar+2sV_x^0\tanh(2\gamma t)\Big)\Big(s\hbar+2V_p^0\tanh(2\gamma t)\Big)-4s{C_{xp}^0}^2\tanh^2(2\gamma t)} \\
&\rightarrow&\frac{s\hbar}{2}\quad\text{ as }\quad t\rightarrow\infty
\nonumber\\
C_{xp}(t)&=&\frac{s\hbar^2C_{xp}^0\sech^2(2\gamma t)}{\Big(\hbar+2sV_x^0\tanh(2\gamma t)\Big)\Big(s\hbar+2V_p^0\tanh(2\gamma t)\Big)-4s{C_{xp}^0}^2\tanh^2(2\gamma t)} \\
&\rightarrow&0\quad\text{ as }\quad t\rightarrow\infty \nonumber
\end{eqnarray}
where $V_x(0)=V_x^0$, $V_p(0)=V_p^0$ and $C_{xp}(0)=C_{xp}^0$. Hence the process of measurement induces the system state to collapse into a coherent state. If the measurement retrieves no information on momentum, i.e. $\Gamma_2=0$, then
\begin{eqnarray}
V_x(t)&=&\frac{\hbar V_x^0}{\hbar+4V_x^0\Gamma_1 t} \\
V_p(t)&=&V_p^0+\hbar\Gamma_1t-\frac{4{C_{xp}^0}^2\Gamma_1 t}{\hbar+4V_x^0\Gamma_1 t} \label{vp}\\
C_{xp}(t)&=&\frac{\hbar C_{xp}^0}{\hbar+4V_x^0\Gamma_1 t}
\end{eqnarray}
and the growth in the momentum variance is unbounded. Similarly, if the measurement retrieves no information on position then the position variance grows unbounded. However, when both position and momentum are measured simultaneously the system state is forced into a coherent state. When $\hat{H}\neq 0$ we expect that for a suitable choice of $\gamma$, any spreading of the quantum wavepacket caused by nonlinearities in the Hamiltonian will be counteracted by the measurement induced localization.

One might naively assume that if the measurement only retrieves information on position (or momentum) then the state will not localize. However this is not always the case. Note that when $|C_{xp}^0|>\hbar/2$ in equation (\ref{vp}) the momentum variance will initially decrease. Thus if the system dynamics is such that it increases the covariance between position and momentum, then the continuous measurement of position may also localize momentum. For example, consider the Hamiltonian describing free particle motion, $\hat{H}=a\hat{p}^2$. When $\Gamma_2=0$ the variances and covariance satisfy
\begin{eqnarray}
\frac{dV_x}{dt}&=&-\frac{4\Gamma_1}{\hbar}{V_x}^2+4aC_{xp} \\
\frac{dV_p}{dt}&=&\Gamma_1\hbar-\frac{4\Gamma_1}{\hbar}{C_{xp}}^2 \\
\frac{dC_{xp}}{dt}&=&-\frac{4\Gamma_1}{\hbar}C_{xp}V_x+2aV_p.
\end{eqnarray}
Although we could not solve these equations analytically, it is easy to see that all physical solutions asymptotically attract to the stable fixed point
\begin{equation}
V_x=\sqrt{\frac{a}{2\Gamma_1}}\hbar, \qquad V_p=\sqrt{\frac{\Gamma_1}{2a}}\hbar, \qquad C_{xp}=\frac{\hbar}{2}.
\end{equation}
Hence, measurement of position does not introduce a diffusion in momentum, and the state localizes (this result has been derived previously in \cite{diosi2}). However, for free particle motion, if we only measure momentum the state does not localize. The system dynamics accelerates the growth in the position variance. See \cite{gisin,gisin2,schack,brun,percival,strunz,strunz2,doherty} for more on localization.

\section{Numerical simulations}
We will now numerically investigate the solution of the stochastic Schr\"{o}dinger equation (\ref{stochschro}) for the driven system
\begin{equation}
H=ap^2+bx^2+cx^4+dx\cos(\omega t).
\end{equation}
The numerical method to solve this equation is simple. To take advantage of the measurement induced localization we use a local moving number basis
\begin{equation}
|n\rangle\equiv\textstyle\frac{1}{\sqrt{n!}}\hat{a}^{\dag n}|0\rangle\qquad n=0,1,\dots N \qquad (s=1).
\end{equation}
truncated at some finite value $N$. The stochastic terms are integrated using the first-order Euler method while other terms are integrated by diagonalizing the position and momentum operators and using the split-operator formula.

We will first consider an integrable case when $a=b=5$, $c=1$, $d=0$ and $\hbar=0.05$. The initial state was chosen to be a coherent state ($s=1$) centered at $(x,p)=(-2,1)$ when $t=0$. The Husimi density of the initial state together with the contours of the Hamiltonian are plotted in Fig. 1(a). The Husimi density of the evolved state ($t=4$) together with the trajectories $(\langle\hat{x}\rangle,\langle\hat{p}\rangle)$ for different measurement schemes is plotted in figures 1(b)-(d). The evolved state in Fig. 1(b) is the result when no measurements occur ($\gamma=0$). In this case, nonlinearities in the Hamiltonian cause the state to shear as it evolves, spreading it along the contours. The trajectory has little meaning when $\gamma=0$. In Fig. 1(c) the evolved state is the result of continuous simultaneous measurement of position and momentum with $\gamma=1/\sqrt{2}$ and $s=1$. In this case the state has remained localized as it follows the contours. The continuous measurement of position only ($\Gamma_2=0$) when $\Gamma_1=1$ has also kept the state localized. This is shown in Fig. 1(d). The combined variance of position and momentum is plotted in Fig. 1(e). We must emphasize that only when both position and momentum are measured together does the trajectory correspond via (\ref{xa},\ref{xb}) to the outcome of an actual measurement. If only position is measured, only $\langle\hat{x}\rangle$ observed while $\langle\hat{p}\rangle$ is simply the result of a mathematical calculation.  

Now consider the chaotic case when $a=5$, $b=-8$, $c=1$, $d=15$, $\omega=2\pi$ and $\hbar=0.05$. A Poincar\'{e} stroboscopic map with unit strobing frequency is plotted in Fig. 2(a). For an initial state the same as above, the evolved state ($t=5$) together with the trajectories for different measurement schemes is plotted in figures 2(b)-(d). When no measurement occurs (Fig. 2(b)) the chaotic action of stretching and folding spreads the state across the phase space. However when we continuously measure position and momentum (Fig. 2(c)) the state remains localized and the trajectory resembles classical motion with noise. This noise will vanish as we approach the classical limit $\hbar\rightarrow 0$. In Fig. 3(a) we have plotted the quantum trajectory for the same parameter values as above except with $\hbar=10^{-6}$. In this case the total variance remains below $10^{-5}$ (Fig. 3(b)) and the noise is not visible. The corresponding classical trajectory is plotted in grey and is only visible when it deviates from the quantum at $t\approx 5$. The evolution of this system under the continuous measurement of position has already been studied by Bhattacharya et al. \cite{habib}. They also find that the measurement keeps the system state localized. It is not surprising that this is also the case when only momentum is measured (Fig. 2(d)).  

\section{Conclusion}

We have derived an It\^{o} stochastic Schr\"{o}dinger equation (\ref{stochschro}) describing the evolution of a quantum system under the continuous simultaneous measurement of position and momentum. The outcome of this measurement is a classical stochastic record obeying (\ref{calX}). As a consequence of the measurement, the system state is forced to remain localized allowing a classical interpretation of the quantum mean of the phase space variables as the trajectory of the system. This trajectory corresponds to the actual measured trajectory minus noise (\ref{xa},\ref{xb}). Furthermore, the localization property allows a well-defined classical limit via Ehrenfest's theorem. Indeed, for small $\hbar$, numerical results show that the quantum system approximately follows classical trajectories. However a more complete theoretical understanding of the classical limit under continuous measurement is needed.

\section{Figure captions}

\noindent
{\bf FIG. 1. (a)} Contours of $H=5p^2+5x^2+x^4$ and the initial state ($\hbar=0.05$). {\bf (b-d)} The trajectory and final state at $t=4$ when (b) $\gamma=0$, (c) $\gamma=1/\sqrt{2}$ and $s=1$, and, (d) $\Gamma_1=1$ and $\Gamma_2=0$. {\bf (e)} The combined variance of position and momentum. All quantities are dimensionless.  \\

\noindent
{\bf FIG. 2. (a)} Poincar\'{e} map for $H=5p^2-8x^2+x^4+15x\cos(2\pi t)$ and the initial state ($\hbar=0.05$). {\bf (b-d)} The trajectory and final state at $t=5$ when (b) $\gamma=0$, (c) $\gamma=1/\sqrt{2}$ and $s=1$, and, (d) $\Gamma_1=0$ and $\Gamma_2=1$. {\bf (e)} The combined variance of position and momentum. \\

\noindent
{\bf FIG. 3. (a)} The quantum (black) and classical (grey) trajectories for the same Hamiltonian as in Fig. 2 except with $\hbar=10^{-6}$. {\bf (b)}The combined variance of position and momentum. \\

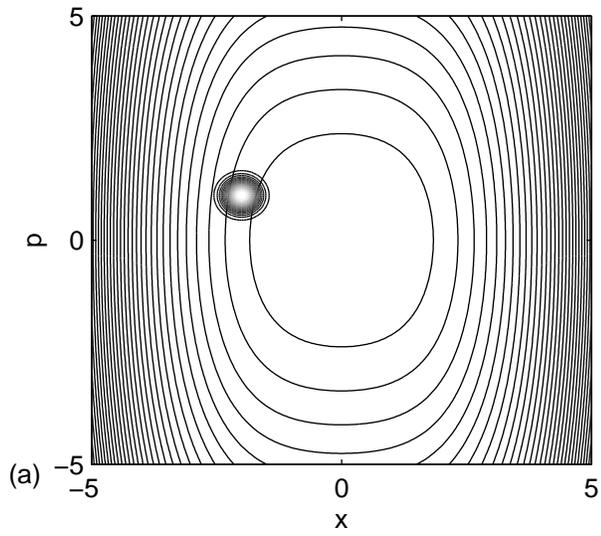
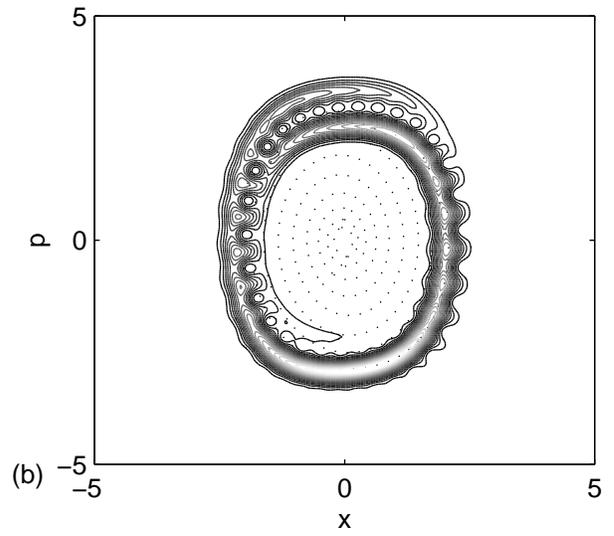
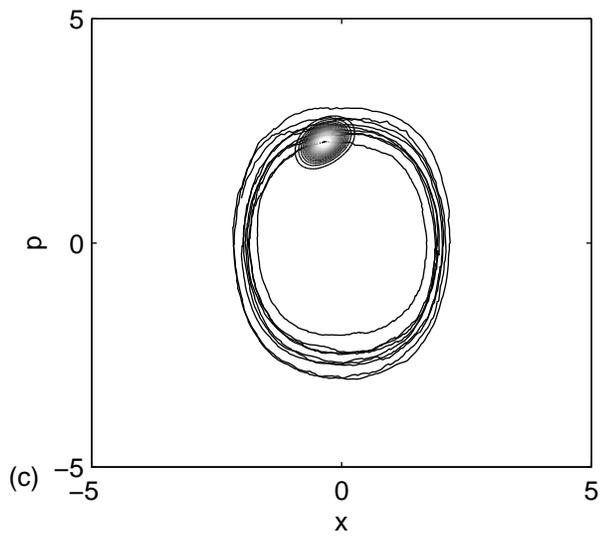
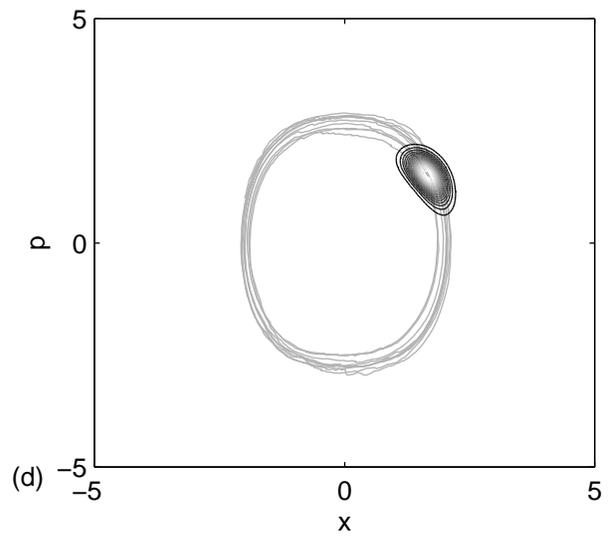
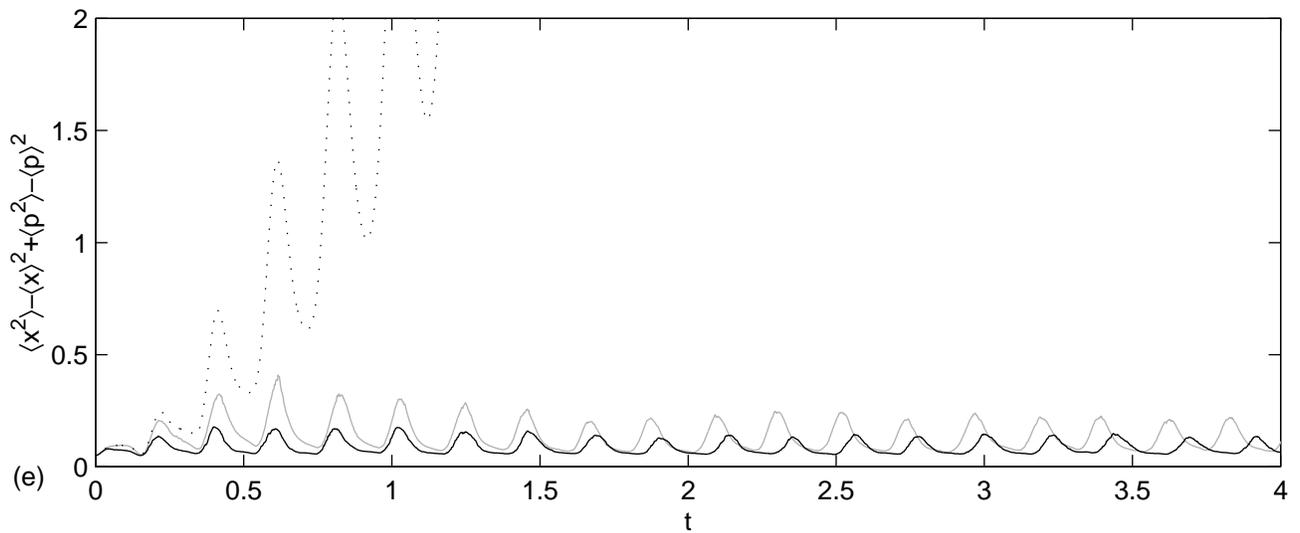
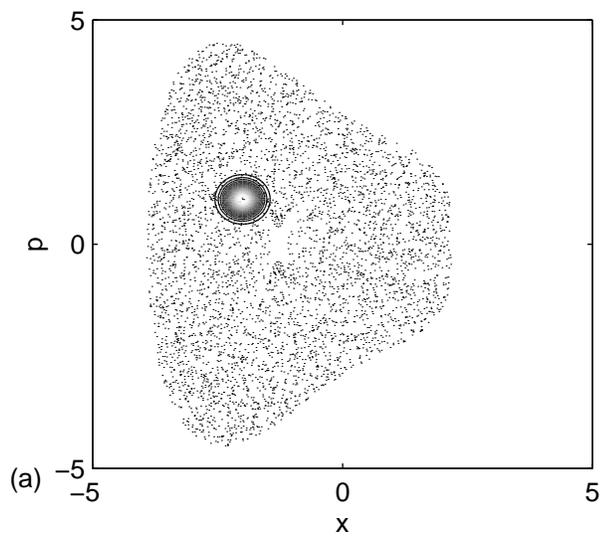
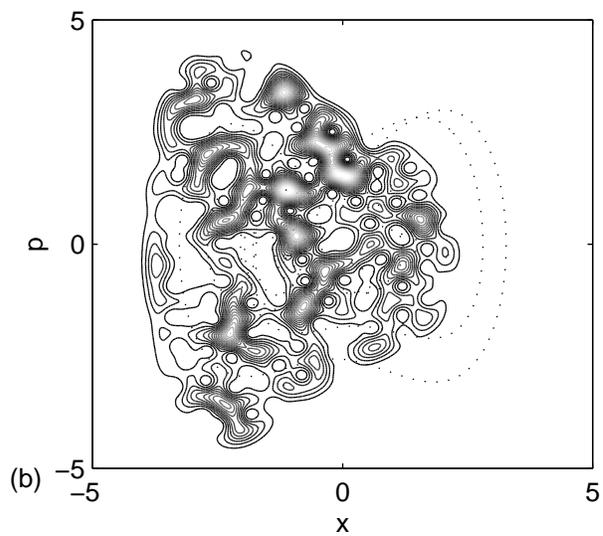
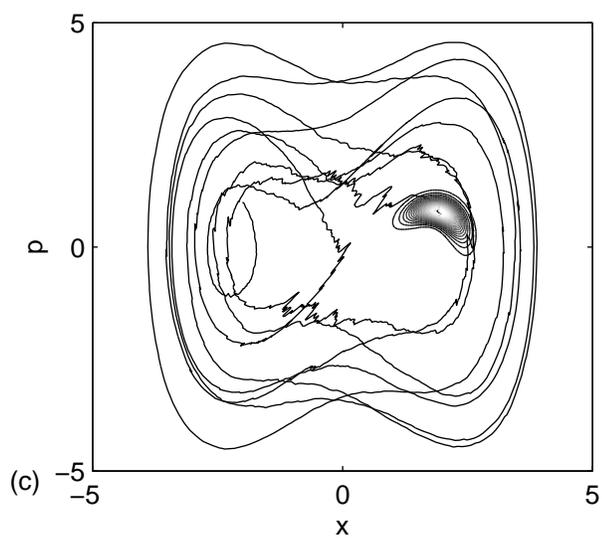
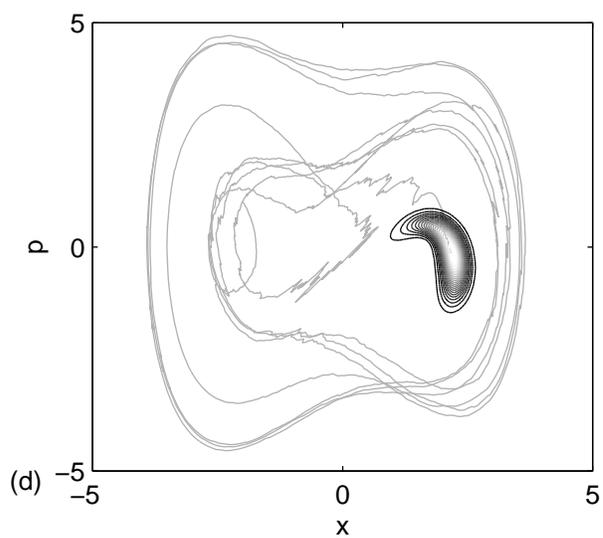
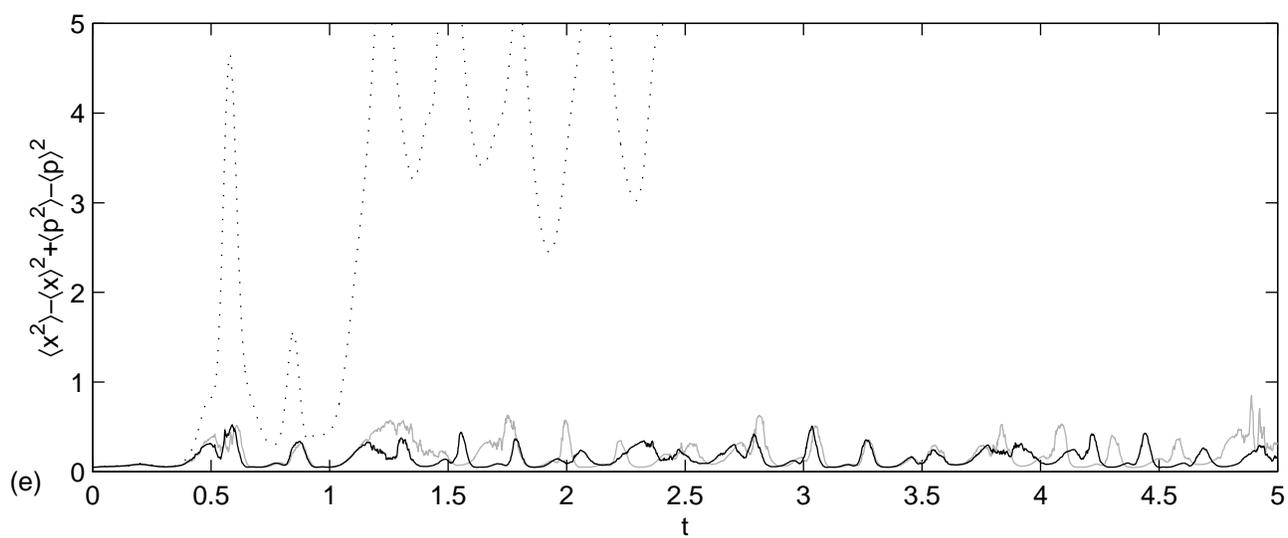
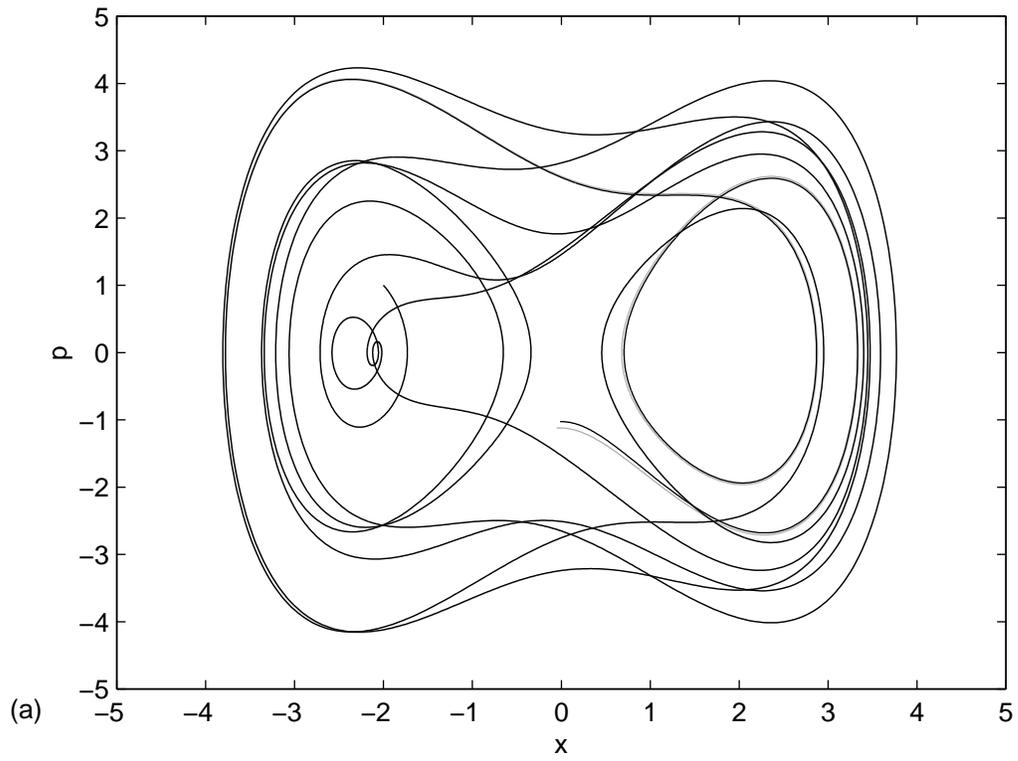
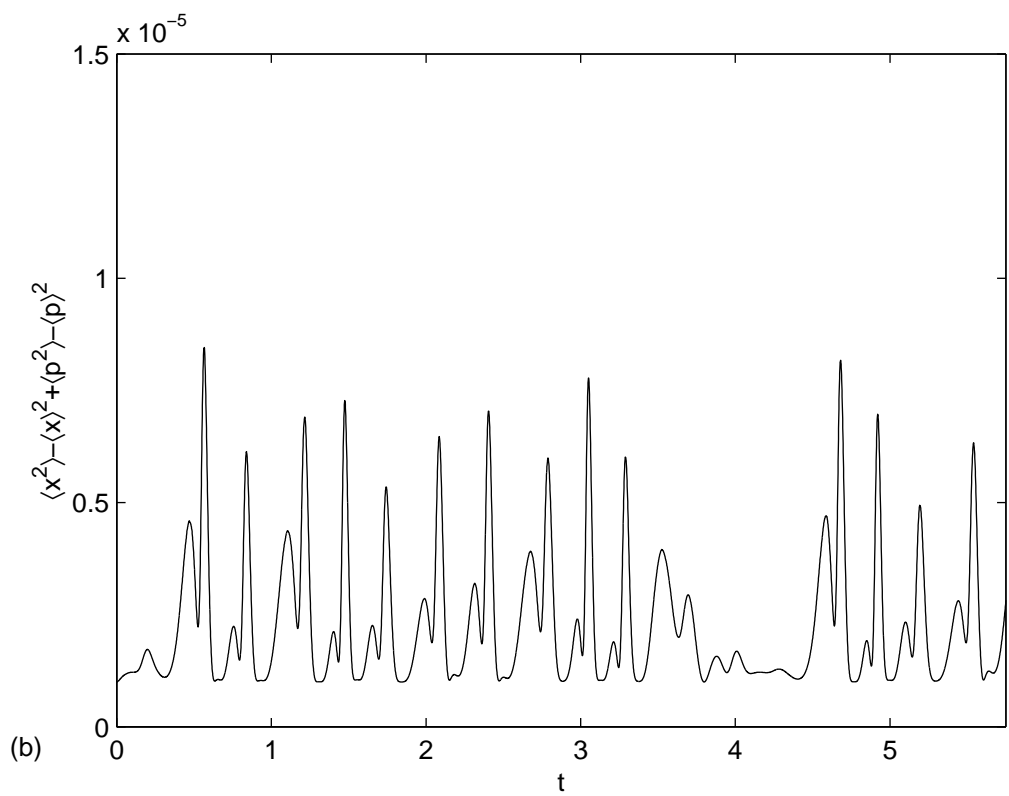
\end{document}